\documentclass[runningheads]{llncs}
\usepackage[T1]{fontenc}
\usepackage{graphicx}
\usepackage{xcolor}
\usepackage[hidelinks]{hyperref}
\newcommand{\approachName}{CBSAE}
\newcommand{\Section}[1]{\textbf{\hyperref[#1]{Section~\ref{#1}}}}

\newcommand{\fig}[1]{\hyperref[#1]{\textbf{Figure \ref{#1}}}}

\begin{document}
\title{Towards a case-based learning approach to support software architecture education}
\titlerunning{Towards a CBL approach to support software architecture education}
%


\author{Brauner R. N. Oliveira\inst{1} \and
Elisa Y. Nakagawa\inst{1}}
%
\authorrunning{Oliveira, B. R. N. and Nakagawa, E. Y.}
%
\institute{¹ University of S\~{a}o Paulo, S\~{a}o Carlos, SP, Brazil\\
\email{brauner@usp.br, elisa@icmc.usp.br}}

\maketitle              
\begin{abstract}
Software architecture education remains challenging for instructors, students, and software industry professionals. Several initiatives have been proposed to mitigate the inherent challenges, including games, supporting tools, collaborative courses, and hands-on projects. Case-based learning has been introduced in software architecture, and its benefits are recognized. However, choosing the right cases that cover the stated learning objectives and developing learning activities to achieve high-order learning are also challenging. The main goal of this paper is to present a case-based learning approach that guides the development of learning objectives, the finding and selection of real-world software architecture cases, and the design of instructional activities. We applied our approach in software architecture related courses during the past few years. The results show that it can leverage the ways to adequately explore cases for educational purposes while also motivating instructors and students to the software architecture education.     

\end{abstract}

\section{Introduction} 

Several initiatives and experiences of teaching software architecture have been published pointing out to what can be considered a consensus: Teaching software architecture is a challenging task \cite{Mannisto2008TSAD,Lago2005TCSA,Rupakheti15TSAU,Galster2016WMTS,Cervantes2016SDAD,Deursen17CATS,Angelov17DAAS,Lago2019DAPC}. Several factors can explain it, including the existing gap between low-level courses (e.g., programming and data structures) and architecture \cite{Garlan1992ECAS}, the difficulty to precisely define what software architectures are exactly due to several existing definitions and notions available\footnote{\href{https://resources.sei.cmu.edu/library/asset-view.cfm?assetid=513807}{What is your definition of software architecture?}}, and the high-order cognitive tasks and knowledge involved in architectural activities \cite{Galster2016WMTS}. Moreover, students have difficulties with solving a given problem when many ways exist to solve that problem, i.e., there is no single best solution \cite{Lago2005TCSA,Galster2016WMTS}; this is the case for software architecture design. In turn, the architectural design process is difficult to be performed with confidence due to many aspects to be considered when making architectural decisions, which are sometimes made along with the software development \cite{Cervantes2016SDAD}. In general, the desired skills of architects are mostly developed or gained through experience over many years designing software architectures \cite{Rupakheti15TSAU,Cervantes2016SDAD,Lago2019DAPC}. Otherwise, instructors have usually a small amount of time during courses to address software architecture topics that students further face in industry and also to teach those topics in an adequate way that ease the path throughout their career. 
Other reasons that make teaching software architecture too difficult are deeply discussed in \cite{Galster2016WMTS}. Without appropriate directions for teaching software architecture, designing suitable architectures for particular problems becomes challenging for software architects.

To overcome the inherent and challenging nature of teaching software architecture, many initiatives were already proposed, as summarized in \cite{Oliveira2022OSAE}. This study reported several initiatives, including types of experience in software architecture education, software architecture topics considered relevant for the learning process, categories of learning objectives, and educational approaches. For instance, \cite{Lago2019DAPC} proposed a card game to support the learning process regarding architectural decision making. Similarly, \cite{Cervantes2016SDAD} developed a card game that leverages the learning of ADD (Attribute-Driven Development). A project-based learning approach was designed and applied to teach architectural design process in agile projects~\cite{Angelov17DAAS}.
Case-based learning has also been explored in different occasions and ways to achieve simple to high-order learning objectives, including in software architecture \cite{Deursen17CATS,Butler1999ClientServerCase,Ouh2019ACBL,Kiwelekar2015SALearningObjectives}. Whereas the use of cases for learning purposes has benefits \cite{Herreid1994CaseStudiesInScience,McLean2016CBLHealthcareReview,Williams2005CBLReviewPreHospital}, the adoption of a case-based approach is not an easy task. Instructors have had diverse inherent difficulties, such as the finding, selection, and development of cases, design of instructional material, design of in-class activities based on cases, and assurance that the learning objectives are achieved appropriately, as also previously presented in \cite{Yadav2007CaseTeachingSurvey}.
Hence, the research question that motivated the conduction of this work was: \textit{How is it possible to adequately use case-based learning for software architecture education?}

The main goal of this paper is to present CBSAE, a case-based learning approach that explores the use of cases for software architecture education. In short, CBSAE encompasses the definition of learning objectives, selection of cases (i.e., finding and assessment of cases as well as conduction of case studies), and design of instructional material and activities. For that, we used Design Science as a research method and conducted three evaluations to verify the relevance of adopting cases and the viability of our approach. As a result, we observe that CBSAE is useful for leveraging software architecture education.

The remainder of this paper is organized as follows. \Section{sec:cbsae} presents CBSAE. \Section{sec:preliminary_evaluation} presents 
results of the preliminary evaluations, whereas \Section {sec:discussion} discusses the results and future work, and \Section{sec:final_remarks} presents the final remarks.

\section{\approachName{}: A Case-Based Learning Approach}\label{sec:cbsae} 

\begin{figure}[ht]
\centering
\includegraphics[width=1\textwidth]{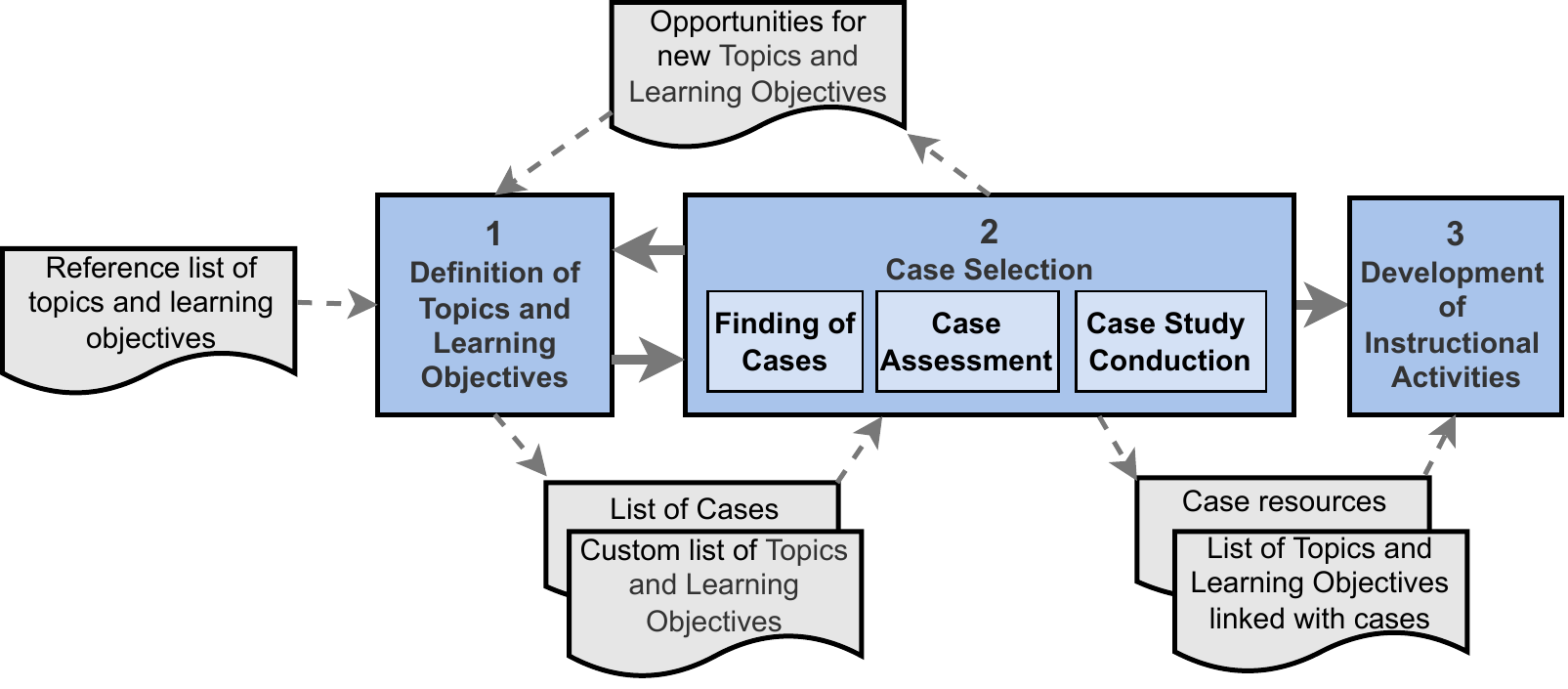}
\caption{Overall View of \approachName{}}
\label{fig:process}
\end{figure}

\approachName{} is composed of three main activities, as shown in \fig{fig:process}. \textbf{1. Definition of Topics and Learning Objectives} aims at establishing the learning goals. \textbf{2. Case Selection} deals with the finding, selection, and/or development of cases and associated materials to covering the topics and learning objectives. \textbf{3. Development of Instructional Activities} establishes how the case(s) will be explored for learning purposes. The first and second activities can feed each other, i.e., topics and learning objectives guide the finding of cases, while cases can be a rich source for discovering new topics and defining learning objectives.

\subsection{Definition of Topics and Learning Objectives}

The topics and learning objectives to be addressed in an educational occasion vary depending on some factors, such as the time available for the occasion, type of occasion (e.g., short or long courses), goals of the instructor(s), institution curricula, among others. Hence, \approachName{} provides a reference list of topics and learning objectives that was defined based on analysis of curricula of several institutions that teach software architecture and also evidence gathered from a systematic mapping study \cite{Oliveira2022OSAE}. In short, this study 
selected and analyzed 50 experiences with software architecture education to understand how the discipline has been taught, the most common topics and learning objectives, and the learning methods and resources used.
Regarding the topics, we managed to find the most common quality attributes, architectural solutions (e.g., patterns and styles), and methods addressed. Each learning objective is categorized regarding the knowledge and cognitive process dimensions, as proposed by the Revised Bloom's Taxonomy (RBT) \cite{Anderson2001RBT}. \textit{``Learn how to apply ATAM to evaluate the architecture of a system''}, for instance, is categorized as \textbf{procedural} in the knowledge dimension (as it requires the student to have knowledge on how to follow method steps) and as \textbf{evaluate} within the cognitive process dimension (as the evaluation process requires participants to make judgments based on criteria). In this activity, instructor(s) may use the reference list (which is not mandatory) to define the topics and learning objectives that will support the next activity and will guide the educational experience. For this, RBT can be employed to define which concepts will be addressed and in which cognitive level. Moreover, the topics and learning objectives can be iteratively refined based on the cases themselves (selected in the next activity).

\subsection{Case Selection}

Considering the preliminary list of topics and learning objectives, the selection of cases suitable to covering that list is performed. To support this activity, \approachName{} provides a conceptual model (referred to as CM4SAC (Conceptual Model for Software Architecture Cases)) with diverse concepts (and relationship among them) associated with software architecture cases. The model includes software architecture core concepts, including architectural views, viewpoints, documentation, and decisions, and case-related concepts, such as contextual, authorial, and temporal information. Hence, CM4SAC can be used for finding out relevant information from cases in the software architecture context. This model also enables instructors to figure out the strengths and weaknesses of a given case, such as which information is missing or information well-detailed with good potential for learning. This activity is composed of two mandatory and one optional sub-activity as follow.

\vspace{.2cm}
\noindent\textbf{Finding of Cases.} Materials addressing software architecture cases can be found in different sources and formats. Books, book chapters, conference proceedings, and journals related to software engineering or architecture, such as ECSA, ICSA, ICSE, and JSS, are relevant sources. Besides scientific literature, gray literature can be considered, such as technical reports from companies and institutions. Development/engineering blogs (or just DevBlogs) are largely used by companies, teams, and individuals to share information on real-world software projects, including information on software architecture. Another important source comes from video streaming platforms such as YouTube, which provide informal or even formal presentations addressing real-world software architecture cases. We have also collected and organized cases that address different topics for software architecture education, such as architectural views, architectural design and evaluation. One of the cases addresses the arXiv-NG project, which aimed to re-architect the system behind \textit{arxiv.org} into a more modern, maintainable, and scalable architecture. Hence, we intend to enrich CBSAE with a set of cases to facilitate this activity.

\vspace{.2cm}
\noindent\textbf{Case Assessment}: 
Once cases are found, the next step involves the assessment of materials and contents to decide whether they are suitable for supporting the development of learning activities to meet the learning objectives. The first concern is to find which information is explicitly available in the cases to enable their complete assessment. The materials associated with a given case may not present information that is relevant for one or more learning objectives. For instance, an instructor may not find it useful to use a case that has no representation of an architecture if the intention is to teach the concept of architectural views, even though that case can match other objectives or serve as basis for the establishment of new ones. To conduct such assessment, CM4SAC can support the assessment of the completeness of cases, as it describes core concepts associated with cases and relationships among concepts.

\vspace{.2cm}
\noindent\textbf{Case Study Conduction}: Some cases can match partially the learning objectives but still are interesting to be used; so additional information or changes are required to complement that cases. For that, \approachName{} provides a method based on \cite{Runeson2012CSRS} to conduct case study covering relevant aspects for this kind of study. This method can also be applied to design cases from scratch, addressing real-world projects with no information published anywhere.

\subsection{Development of Instructional Activities}

This step matches the list of topics and learning objectives with the cases for developing instructional activities. 
A number of activities can be developed based on cases, for example, in-class discussions, debates, story-driven lessons, and theoretical and practical assignments. An activity in which students perform the analysis of a microservice architecture is an example that matches the learning objective \textit{``Learn how the microservice architecture style organizes system components''} with a case that employs that style. \approachName{} provides guidelines to support instructors in developing those activities and also presents actual examples based on cases (which we have been using).


\section{Preliminary Evaluation}\label{sec:preliminary_evaluation} 

To evaluate our approach, we performed three preliminary evaluations in the context of computing-related undergraduate courses at our university. The first two evaluations assessed whether the use of software architecture cases developed using our approach are beneficial for learning from the perspective of students. The third evaluation aimed at assessing students perspective on learning using the cases and collecting qualitative feedback to improve our approach. These evaluations are detailed in the following.

The first evaluation was conducted in the first semester of 2019. After reading the material on arXiv-NG project (mentioned before) that consisted of architectural documentation and a story-telling text, the students solved questions that addressed architectural concepts at different cognitive levels. Following, we applied a questionnaire to assess their perception of using the case for learning activity besides the difficulties faced. Optionally, they provided qualitative feedback on any aspect of the activity. In total, 53 students did the exercise and anonymously and individually answered the questionnaire.
The second evaluation was similar to the first one, except that we (the instructors) presented and discussed the arXiv-NG case during class time. We also addressed questions from students and discussed different aspects of the case and software architecture. Later, the students answered questions similar to those applied during the first evaluation. Following, they responded to a questionnaire that assessed their perception on the use of cases for learning software architecture. This evaluation was performed in the second semester of 2020 in three different student classes, and a total of 164 
students answered the questionnaire anonymously and individually.

\begin{figure}[h]
\centering
\includegraphics[width=1\textwidth]{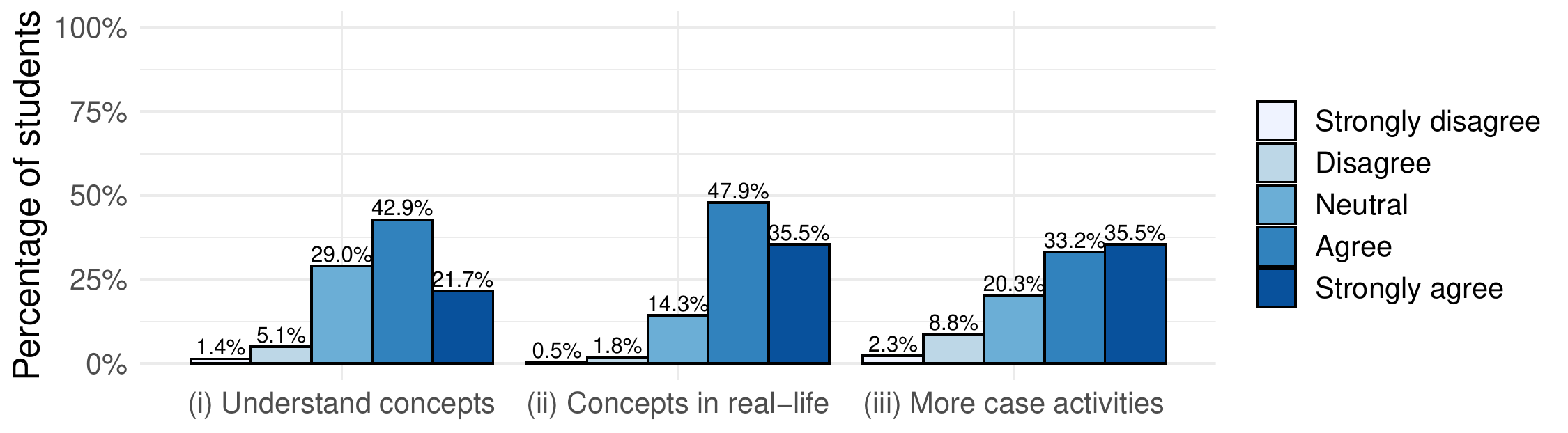}
\caption{Results of the first and second evaluations}
\label{fig:q_results_case_perception}
\end{figure}

In both evaluations, we asked students to evaluate three premises: (i) \textit{``The case was useful to help me understand concepts associated with software architecture''}; (ii) \textit{``The case was useful to help me understand how concepts related to software architecture are present in real-world software projects''}; and (iii) \textit{``I would like to do more activities/exercises using software architecture cases''}. \fig{fig:q_results_case_perception} shows the results (collected using Likert scale from ``Strongly disagree'' to ``Strongly agree''). As observed, most students agree or strongly agree that software architecture cases are relevant to learning purposes. Hence, we believe the use of \approachName{} to organize, write cases, and develop instructional activities can leverage the learning process for software architecture education experiences.

The third evaluation occurred in the second semester of 2021. We followed a more case-oriented way in which different software architecture cases were applied throughout the semester. These cases were mainly used to explore different cognitive levels. We followed \approachName{} to define topics and learning objectives, select cases, and develop instructional activities. We also asked the students to search for cases that they considered interesting and relevant to be used throughout the course and for a final project (which involved the architectural design of a system using C4 and evaluation with ATAM).
Following, to gather qualitative feedback, we performed semi-structured interviews after the end of the semester with eight volunteer students and addressed three main issues: the cases that we selected, the cases selected by the students, and the application of \approachName. It resulted in a total of 4 hours and 14 minutes of interviews that were recorded to further deeper analysis. In short, we asked questions to gather their overall opinion on the contribution of cases for their learning besides the way the cases were employed. For instance, we assessed whether they preferred to have a case presented to them or deep dive into the case material to learn from it. 
As the main results regarding the cases (which we selected), students had a positive learning experience in terms of achieving learning objectives. From their point of view, the availability of different materials (e.g., video and text) contributed to better understanding, solving exercises, and developing in-class discussions. Regarding the cases selected by them, students believed that some cases were not entirely useful due to the lack of documentation. In both situations (cases selected by us vs. cases selected by them), they reported that familiarity with the cases under consideration was relevant for motivation and learning. In other words, cases that employed technologies, frameworks, and languages, for instance, were considered more pertinent to address architectural concepts. Moreover, some students preferred to learn using the cases they selected, whereas others preferred to use the cases we presented.

\section{Discussions}\label{sec:discussion} 


 
The idea to develop \approachName{} was based on the benefits of using cases highlighted in several studies, even in other knowledge areas, such as science, medicine, and law. In our preliminary evaluation, almost all students recognized the value of cases for learning software architecture; this result is also aligned with the literature and practice in software architecture education. As stated before, however, applying a case-based learning approach imposes several challenges for instructors. As a result, \approachName{} addresses cases as central sources for learning and intends to support educators in different learning scenarios, ranging from short learning sessions (which aim at employing real-world cases, like lectures and workshops) to semester or year-long courses of software architecture. The latter is the primary goal of \approachName{} by making it possible to explore several cases within the time available. 

For future work, we intend to conduct evaluations considering the instructors' point of view when using \approachName{}, its activities, and artifacts. Another future work is to provide support to the assessment of students' learning, as such assessment is considered challenging when using cases.
 
\section{Final Remarks}\label{sec:final_remarks} 
It is well-known that case-based learning has its strengths and weaknesses. This paper presented \approachName{}, an approach that takes advantage of such strengths while mitigates the weaknesses. The results achieved so far are very promising and motivate us to improve and refine \approachName{} and, ultimately, it can contribute to leveraging the education in software architecture.


\bibliographystyle{unsrt}
\bibliography{references}

\end{document}